After the LaTeX file is an uuencoded, compressed, tar-archive of postscript figures.

\documentstyle[12pt,epsf]{article}
\begin{document}
\newcommand{\be}{\begin{equation}}
\newcommand{\ee}{\end{equation}}
\newcommand{\bea}{\begin{eqnarray}}
\newcommand{\eea}{\end{eqnarray}}
\newcommand{\half}{\frac12}
\newcommand{\eq}[1]{eq.~(\ref{#1})}
\newcommand{\daw}{{\cal D}}
\newcommand{\art}[6]{\bibitem{#1}#2, #3 #4 (#5) #6.}

\title{Monte Carlo calculation for systems consisting of several
coordinate patches}
\date{October 17, 1994}
\author{Claus Vohwinkel\\
{\small{\it
Supercomputer Computations Research Institute, The Florida State
University,}} \\
{\small{\it
Tallahassee, FL 32306, USA}}}
\maketitle

\begin{abstract}
I investigate the time step dependence of Monte Carlo simulations for
coordinate-spaces consisting of several patches. It is shown that a
naive kinetic term does not necessarily converge to the 
same spectrum as a Hamiltonian calculation. Then an improved kinetic
term is presented which allows one to connect the Monte Carlo and
Rayleigh-Ritz results of intermediate volume SU(2) gauge theory.
\end{abstract}

\section{Introduction}
There have been two approaches in the intermediate volume calculation
of SU(2) gauge theory: The Hamiltonian approach \cite{baal},
which is a Rayleigh-Ritz analysis of an effective Hamiltonian.
and the Lagrangian approach~\cite{mic0}\cite{mic} which is a Monte
Carlo simulation of an effective Lagrangian.
For a review of most of the work done in this area see~\cite{baal3}.
Although both approaches are formally equivalent~\cite{baal2},
the numerical data obtained does not agree. A study of
the Lagrangian approach using smaller time steps did not improve the
results \cite{tiede}.

In both approaches the coordinate space consists of 8 compact patches,
which correspond to equivalent vacua. Basically these patches are
spheres
and tunneling from one sphere into another one can take place via any point
of the boundary of the sphere. This means, 
that in flat space, we cannot form a single large patch
by joining all the patches together.
The idea, that this kind of geometry is responsible
for the difference of the Lagrangian and Hamiltonian results, is very tempting.
Indeed, the disagreement in the energy spectrum is still present,
when one switches off the potential energy. One can
therefore study the simpler case of a free particle moving in the
same geometry, which allows one to use analytical methods.
This is carried out in section 2.
After having understood the reason for the disagreement,
one can change the kinetic term in the Lagrangian approach
to match the Hamiltonian approach. A procedure to do so and some
Monte Carlo results are given in section 3. Finally section 4 contains
some conclusions.

\section{A toy model}
The model without the potential energy factorizes
into 3 wave functions. For the analysis attempted here,
it is sufficient to describe one of these wave functions:
Our toy model consists of a free particle moving in two spheres, each of
radius $\pi$. The obvious choice of coordinates are
spherical coordinates and an integer $k=0,1$, giving the sphere in
which the particle is:
\be \tilde{\psi} = \tilde{\psi}(r,\theta,\varphi,k). \ee
The spheres touch each other at their boundary.
(One can picture this a two spheres occupying the
same space.) For the
particle to move from the interior of one sphere to the other one, it has to move to any
point on the boundary $r=\pi$ where it changes to the other sphere and
then moves on inward.

Due to the symmetry of the problem the eigenfunctions
will be either symmetric or anti-symmetric under interchange of the
spheres. One can therefore describe the wave function by its wave
function $\psi$ in one sphere and its symmetry type.

For the anti-symmetric wave functions the boundary condition at
$r=\pi$
is unambiguous:
\be \psi(r,\theta,\varphi) = 0. \label{asy} \ee
For the symmetric wave function the choice is not as clear cut. Due to
the measure one
should however look at $r\psi(r,\theta,\varphi)$ instead of
$\psi(r,\theta,\varphi)$.
One can select a basis $B(b)$ by requiring
\be \left. \frac{\partial}{\partial r} \log( r\psi(r,\theta,\varphi)) \right|_{r=\pi}
= b, \label{sy}\ee
for a given $b$.
In the original problem the angular variables could not be observed, and
the obvious choice was $b=0$. Here, however, one is free to choose any $b$.
The eigenfunctions are given by
\be \psi_{ilm}(r,\theta,\varphi) = j_l(\alpha_i r) Y_{lm}(\theta,\varphi) \label{ef}\ee
where $\alpha_{i-1}$ is the i'th zero (\ref{asy}) or (\ref{sy}). That
these functions form a basis for any $b$, follows from the orthogonality properties
of the Bessel functions.

In the Lagrangian approach, or to be more precise
the Monte Carlo calculation with finite time step $\epsilon$ \cite{mic}, the configurations 
are
described by a
spatial coordinates $\vec{r}(t)$ and a coordinate $k(t)\in\{0,1\}$ which
describes in which sphere the particle is. The measure is flat in
$\vec{r}$ and one possible kinetic term is given by~\cite{baal2}:
\be T_{kin}(\vec{r}(t),\vec{r}(t+\epsilon),\Delta k) = \frac{1}{2\epsilon^2} 
(\vec{r}(t)-\vec{r}(t+\epsilon))^2
+\Delta k
\frac{(\pi^2-r^2(t))(\pi^2-r^2(t+\epsilon))}{2\pi^2\epsilon^2},\label{kin1} \ee
where $\Delta k = |k(t)-k(t+\epsilon)|$.
In \cite{mic} it was proposed that, for $\Delta k = 1$ one should calculate
the geodesic distance $d_g$ between $c(t)$ and $c(t+\epsilon)$ and
use
\be T_{gd} = \frac{1}{2\epsilon^2} d_g^2.\label{kin2}\ee
It was argued that both kinetic terms should give the same continuum
limit, but~(\ref{kin2}) should have smaller finite $\epsilon$ effects.
Unfortunately the geodesic distance is the root of a quartic equation
which makes analytic calculations hard. I shall therefore restrict a
transfer matrix analysis to the Lagrangian~(\ref{kin1}). Later on I
shall do a numerical comparison with~(\ref{kin2}).

\subsection{Transfer matrix calculations}
One can calculate the energy spectrum for finite time step $\epsilon$
by means of the transfer matrix approach. To this end let me
calculate the
transfer matrix between two wave functions $\psi$ and $\phi$
\be T_{\tilde{\psi},\tilde{\phi}}=\sum_{k_1,k_2=0}^1 \int d\vec{r_1}\, d\vec{r_2}\,
\tilde{\psi}(\vec{r_1},k_1)\tilde{\phi}^*(\vec{r_2},k_2)K(\vec{r_1},\vec{r_2},|k_1-k_2|),\ee
where
\be K(\vec{r}_1,\vec{r}_2,\Delta k) = \frac{1}{\sqrt{2\pi \epsilon}^3}
\exp\left[-\epsilon T_{kin}(\vec{r}_1,\vec{r}_2,\Delta k)\right].\ee
Due to the symmetry we can split up the transfer matrix into and
interior ($T_0$) and exterior ($T_1$) part:
\be
T(\tilde{\psi},\tilde{\phi}) = \left\{ \begin{array}{ll}
T^0(\psi,\phi)+T^1(\psi,\phi) & ,\,\psi\,\rm{and}\,\phi\,\rm{symmetric}\\
T^0(\psi,\phi)-T^1(\psi,\phi) & ,\,\psi\, \rm{and}\,\phi\,\rm{antisymmetric}\\
0&,\,\rm{otherwise}\end{array}\right.\ee
with
\be T^k(\psi,\phi) = \int d\vec{r_1}\, d\vec{r_2}\,
\psi(\vec{r_1})\phi^*(\vec{r_2})K(\vec{r_1},\vec{r_2},k),\ee
where $\psi$ and $\phi$ are wave functions of the form~(\ref{ef}).
Due to the spherical symmetry of the problem, wave functions with
different angular momentum do not mix and one has
\be
T^k(\psi_{ilm},\phi_{jpq}) = T^k_{ij,l} \delta_{lp}\delta_{mq}.\ee

\subsection{Expansion for small time steps}

The matrix elements can be calculated as a series in $\epsilon^{1/2}$. I shall restrict
myself here to the more interesting case of the symmetric sector.
Also I shall give results for $l=0$ only and drop the
$l$ and $m$ references and angular variables from the notation.
One obtains for the diagonal elements:
\be T_{ii} = e^{-\frac12 a_i^2\epsilon}\left[ 1+
 	\frac{a_i^2\epsilon}{12\pi(\pi b^2-b+\pi a_i^2)}(a_i^4\epsilon-6a_i^2\epsilon-12\pi
b+6)\right]+O(\epsilon^{3/2}).\ee
(Note that $a_i$ can be arbitrary large, that is $a_i^2\epsilon$ is not
necessary small.)
For the off-diagonal elements, $a_i$ of order 1 and $a_j$ arbitrary, one has
\be
T_{ij} = -N_iN_j(-1)^{i-j}\left(\frac{2\pi b+1}{\pi^2a_j^2}+\frac{a_j^2\epsilon+2\pi 
b+1}{\pi^2
a_j^2}e^{-\half a_j^2\epsilon}\right) + O(\epsilon^{3/2}),\ee 
with
\be N_i=\sqrt{\frac{\pi a_i}{\pi b^2-b+\pi a_i^2}}.\ee

For $a_j$ of order 1 one will need the next order in $\epsilon^{1/2}$ as
well, and the matrix element for both, $a_i$ and $a_j$ of order 1 is given by:
\be
T_{ij} =
-(-1)^{i-j}N_iN_j\left(\frac{(2\pi b-1)\epsilon}{2\pi^2}+\frac{\sqrt{8}(\pi
b-1)^2\epsilon^{3/2}}{3\pi^{7/2}}\right)+O(\epsilon^2).\ee

Let us assume we can do perturbation theory around
$\Psi = \psi_{i}(r)$.
The first order corrections to the groundstate wave function are
\be \langle \psi_j \Psi \rangle = \frac{T_{ji}}{E_i-E_j}.\ee
For small $a_j$ one has
\be \langle \psi_j \Psi \rangle = (2\pi-b)O(1)+O(\epsilon^{1/2}).\label{sam}\ee
and for $a_j \geq O(1/\sqrt{\epsilon})$ one has
\be \langle \psi_j \Psi \rangle_l = -\frac{(-1)^{i-j}a_0}{\pi^{3/2}\sqrt{\pi b^2-b+\pi a_0^2}}
\frac{(2\pi b+1)e^{\half a_j^2\epsilon}-a_j^2\epsilon-2\pi b -1}
{a_j^2\left(e^{\half a_j^2\epsilon}-1\right)}\label{lam}\ee
Note that for
$b=-\frac{1}{2\pi}$ the corrections~(\ref{lam}) to the wave function
vanish exponentially, once $ a_j^2 > \epsilon$, whereas for all other
choices for $b$ they only vanish as $a_j^{-2}$.
Thus, for any finite $\epsilon$ we expect the correct wave function to obey boundary
conditions~(\ref{sy}) with $b = -\frac{1}{2\pi}$, but the
region where the slope of the wave function becomes compatible with this
b.c. becomes more narrow as $\epsilon$ decreases.
In this simple argument I ignored the fact that in the basis
$b=-\frac{1}{2\pi}$ the first $O(1/\sqrt{\epsilon})$ trial functions contribute
substantially to the true groundstate wave function. Later on I shall
give a more rigorous derivation of the boundary condition.

From \eq{sam} one sees, that the only basis
were one is allowed to do perturbation theory (in $\epsilon^{1/2}$) is
$b=\frac{1}{2\pi}$.

For this choice of $b$ the matrix element~(\ref{lam}) reads:
\be \langle \psi_j \Psi \rangle_l = -\frac{2a_i(-1)^{i-j}}{\pi\sqrt{4\pi a_i^2-1}}
\frac{ 2e^{\half
a_j^2\epsilon}-a_j^2\epsilon-2}{a_j^2\left(e^{\half a_j^2\epsilon}-1\right)}\label{lxx}\ee
For $a_j < O(\epsilon^{-1/4})$ the dominant contribution comes from
the $O(\epsilon^{1/2})$ term in~(\ref{sam}):
\be \langle \psi_j \Psi \rangle_s = \frac{4\sqrt{2}(-1)^{i-j}a_ia_j\epsilon^{1/2}}
{3\pi^{3/2}(a_j^2-a_i^2)\sqrt{(4\pi^2a_i^2-1)(4\pi^2a_j^2-1)}}
.\label{sxx}\ee
Let me calculate the sum over all matrix element squared to see whether
perturbation theory can be applied to this basis.  From \eq{sy} one has
$a_j - a_{j-1} \approx 1$, and because we are only 
interested in the order of the correction, we can replace the sum by 
an integral. Eq.~(\ref{lxx}) gives a contribution
\be
 \sum_{j\neq i} \langle \psi_j \Psi \rangle \approx \epsilon^{3/2}
\int_0^\infty da\,
  \frac{2e^{\half a^2}-a^2-2}{a^2(e^{\half a^2}-1)} = O(\epsilon^{3/2}), \ee
whereas \eq{sxx} contributes
\be
 \sum_{j\neq i} \langle \psi_j \Psi \rangle_s \approx
  \int_1^\infty da \frac{\epsilon}{a^2} = O(\epsilon).
\ee
Thus, corrections to the groundstate wave function vanish as
$\epsilon\rightarrow 0$, and one can indeed do perturbation theory
about $\epsilon=0$.

One is now in a position to determine the energy spectrum $E_i =
-\log(T_{ii})/\epsilon$ of the Lagrangian.
I am using the basis $b=\frac{1}{2\pi}$ and restrict myself to the
low-lying states $a_i = O(1)$.

In addition to the diagonal element $T_{ii}$
one needs the second order correction 
\be \Delta V_i = \sum_{j\neq i} \Delta V_{ij} = \sum_{j\neq i}  
\frac{T_{ij}T_{ji}}{T_{ii}-T_{jj}}.\ee
The only term contributing to $O(\epsilon^{3/2})$ is
\be \Delta V_{ij} = \frac{4a_i^2\epsilon^2}{\pi^2(4\pi^2a_i^2-1)}
\frac{\left((2+a_j^2\epsilon)e^{-\half a_j^2\epsilon}-2\right)^2}
{a_j^4\epsilon^2(1-e^{-\half a_j^2\epsilon})}\label{del}\ee
For small $j$ the contribution $\Delta V_{ij}$ from~(\ref{del}) vanish as $\epsilon^3$ and can 
be
neglected. Therefore one can replace the sum over $j$ by and integral
over $a$ to obtain
\be \Delta V_i = \frac{4a_i^2\epsilon^{3/2}}{\pi^2(4\pi^2a_i^2-1)}
\int_0^\infty da\,\frac{\left((2+a^2)e^{-\half a^2}-2\right)^2}{a^4(1-e^{-\half a^2})}.\ee
Together with the diagonal element $T_{ii}$ up to $O(\epsilon^{3/2})$
\be
T_{ii} = e^{-\half a_i^2 \epsilon} + \frac{4 a_i^2
\epsilon^{3/2}}{3\sqrt{2}\pi^{3/2}(4\pi^2a_i^2-1)}+O(\epsilon^2),\ee
one obtains for the energy
\be E_i = \half a_i^2 - 0.274112 \frac{a_i^2
\sqrt{\epsilon}}{4\pi^2a_i^2-1}.\ee
Eq.~(\ref{sy}) gives $a_0 = 0.3710096482\ldots$,
and one gets for the groundstate energy:
\be E_0 = 0.06882408 - 0.0085092 \sqrt{\epsilon}.\ee

Using the same approach in the anti-symmetric sector 
one obtains for the energy of the lowest state
\be E_a = \half + \frac{1}{2\pi^2}\epsilon.\ee

Let me come back to the boundary conditions in the symmetric sector.
We can confirm our previous analysis about the derivative at the boundary by calculating it in 
the
basis $b=\frac{1}{2\pi}$.
To this end let me write
\be\frac{\partial}{\partial r}\log(r\psi(r,\theta,\phi)) =
 \lim_{u\rightarrow 0} \frac{\Phi(\pi)-\Phi(\pi-u\sqrt{\epsilon})}{u\sqrt{t}\Phi(\pi)},\ee
where $\Phi(r) = r\Psi_i(r,\theta,\phi)$.
Let me restrict myself to the case $l=0$.
Except for $\psi_0$ the only relevant contribution come from large $j$,
and one has:
\be
\Phi(\pi)-\Phi(\pi-u\sqrt{\epsilon})
=\frac{\sqrt{8}a_i\sqrt{\epsilon}} {\pi^{3/2}\sqrt{4\pi^2a_i^2-1}}
 \left[ -\frac{u\pi}{2}
  -\int_0^\infty da\, \frac{\cos(au)-1}{e^{\half a^2}-1} \right] \label{bca} \ee
The integral in \eq{bca} is of order $u^2$ and its value can be ignored.
One obtains for derivative
\be\frac{\Phi(\pi)-\Phi(\pi-u\sqrt{\epsilon})}{u\sqrt{\epsilon}\Phi(\pi)}=-\frac{1}{2\pi}+O(u)
.\ee
Thus we arrive the previous result. Because the first correction in $u$
is $O(u)$ (as opposed to $O(u\epsilon^{1/2})$), one has
$\Phi^\prime(r)/\Phi(r) \approx -\frac{1}{2\pi}$ for $|r-\pi| <
\sqrt{\epsilon}$. But further away from the boundary and small enough
$\epsilon$ the wave function becomes compatible with the lowest
function of the $b=\frac{1}{2\pi}$ basis.

\subsection{Rayleigh-Ritz and Monte Carlo results}
Instead of expanding the matrix elements $T_{ij}$ in a series in
$\epsilon^{1/2}$, they can be calculated numerically 
(ignoring terms of order $\exp[-2\pi^2/\epsilon]$)
and used for a Rayleigh-Ritz analysis. 
The internal elements $T^0_{ij}$ can be integrated
analytically, while the external elements $T^1_{ij}$ can be reduced to a one
dimensional integral which has to be performed numerically. Due to
the exponential fall-off of the matrix elements $T_{ij}$ for large $j$,
the basis $b=-\frac{1}{2\pi}$ is best suited for this
calculation. Depending on $\epsilon$, between 50 and 800 basis states
were used. For the three largest $\epsilon$, the transfer matrix in the
$l=1$ sector was calculated by means of a two dimensional numerical integral.
The results for the vacuum $E_0$, the lowest anti-symmetric state $E_a$,
and the lowest $l=1$ state $E_1$ are given in table~\ref{trr}.
\begin{table}
\begin{center}
\caption{Rayleigh-Ritz results. Shown are the number of basis states
$n_b$ used in the Rayleigh-Ritz calculation, the groundstate energy
$E_0$, the lowest state in the anti-symmetric sector $E_a$ and the
lowest $l=1$ state $E_1$ . All $E_1$ estimates were computed with 50
basis states.}\label{trr} 
\vspace{.25cm}
\begin{tabular}{|l|l|l|l|l|}
\hline
$\epsilon$ & $n_b$ & $E_0(\epsilon)$  &$E_a(\epsilon)-\half$ & $E_1(\epsilon)$ \\ \hline
 $2.0\cdot 10^{-1}$&$ 50 $&$ 0.066552  $&$ 1.27270\cdot 10^{-2}$ &$0.308909$\\
 $1.0\cdot 10^{-1}$&$ 50 $&$ 0.066858  $&$ 5.92709\cdot 10^{-3}$ &$0.306617$\\
 $5.0\cdot 10^{-2}$&$ 200 $&$ 0.067272 $&$ 2.82432\cdot 10^{-3}$ &$0.305885$\\
 $2.0\cdot 10^{-2}$&$ 300 $&$ 0.067757 $&$ 1.08418\cdot 10^{-3}$ & \\
 $5.0\cdot 10^{-3}$&$ 300 $&$ 0.068256 $&$ 2.61903\cdot 10^{-4}$ & \\
 $2.0\cdot 10^{-3}$&$ 400 $&$ 0.068457 $&$ 1.03473\cdot 10^{-4}$ & \\
 $2.0\cdot 10^{-4}$&$ 600 $&$ 0.068706 $&$ 1.01993\cdot 10^{-5}$ & \\
 $2.0\cdot 10^{-5}$&$ 800 $&$ 0.068786 $&$ 1.01536\cdot 10^{-5}$ & \\
\hline
\end{tabular}
\end{center}
\end{table}
From the results one can
estimate the next term in the expansion of $E_0(\epsilon)$ in term of $\epsilon$:
\be E_0(\epsilon) \approx 0.06882408-0.0085092\sqrt{\epsilon}+0.007\epsilon\label{hiher}\ee
In figure~\ref{vacuum} the Rayleigh-Ritz results as well as the two
series are shown.
\setlength{\unitlength}{1cm}
\begin{figure}
\begin{picture}(10,7)
\put(1.8,6.5){$-\Delta E_0$}
\put(7.4,-0.2){$\epsilon$}
\epsfxsize=10cm
\centerline{\epsfbox{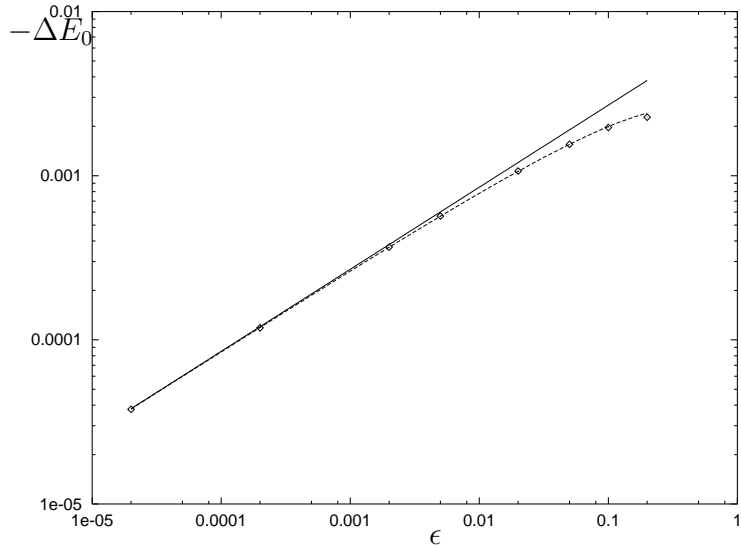}}
\end{picture}
\caption{Comparison of the Rayleigh-Ritz calculation (diamonds),
expansion up to $\epsilon^{1/2}$ (solid line) and estimated expansion up
to $O(\epsilon)$ (dashed line) for the groundstate energy $\Delta E_0 =
E_0(\epsilon)-E_0(0)$. \label{vacuum}} 
\end{figure}

Using the data from table~\ref{trr} one can calculate the mass gaps
for the first antisymmetric state $m_a$ and the first $l=1$ state
$m_1$.
 I have also calculated the mass gaps for these sectors by means of Monte
Carlo simulation, using both kinetic terms,~(\ref{kin1}) and~(\ref{kin2}).
Table~\ref{www} shows the results.

\begin{table}
\begin{center}
\caption{Monte Carlo results for the toy model. The table shows
Rayleigh-Ritz and Monte Carlo results for $m_a$ and $m_1$. $RR$ denotes
the Rayleigh-Ritz calculation, $MC$ the Monte carlo results obtained by
using the kinetic term~(\protect\ref{kin1}) and $gd$ are the masses for the
simulation with kinetic term~(\protect\ref{kin2}).}\label{www} 
\vspace{.25cm}
\begin{tabular}{|l||l|l|l||l|l|l|}
\hline
$\epsilon$ & \multicolumn{3}{c||}{$m_a$} & \multicolumn{3}{c|}{$m_1$}
\\ \hline
 & \multicolumn{1}{c|}{$RR$} & \multicolumn{1}{c|}{$MC$} & \multicolumn{1}{c||}{$gd$}
 & \multicolumn{1}{c|}{$RR$} & \multicolumn{1}{c|}{$MC$} & \multicolumn{1}{c|}{$gd$} \\ \hline
0.05 &  0.435553 &  0.436(5) & 0.336(4) & 0.23862 & 0.2378(15) &  0.2672(10)
\\ \hline
0.10 &  0.439069 &  0.440(4) & 0.338(2) & 0.23976 & 0.2394(10) &  0.2696(10)
\\ \hline
0.20 &  0.446175 &  0.446(2) & 0.341(2) & 0.24236 & 0.2420(4) &  0.2668(6)
\\ \hline 
\end{tabular}
\end{center}
\end{table}

\begin{figure}
\epsfxsize=10cm
\centerline{\epsfbox{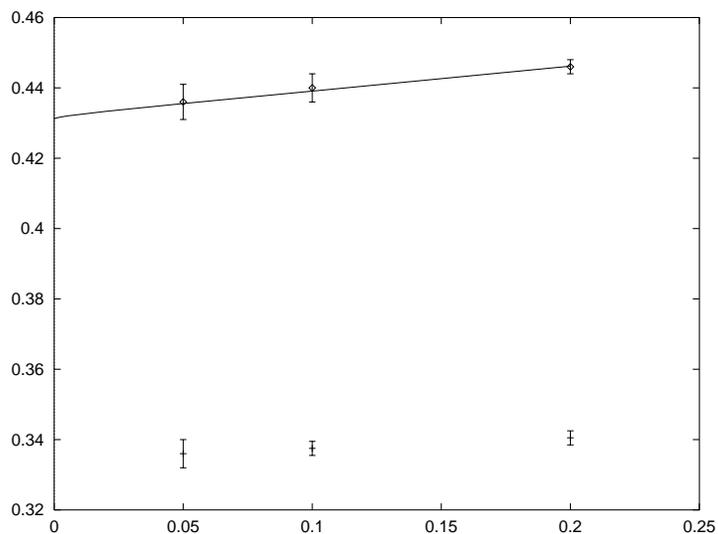}}
\caption{Results for $m_a$. The solid line is the Rayleigh-Ritz
calculation, the diamonds are the Monte Carlo simulation using the
kinetic term~(\protect\ref{kin1}) and the crosses are
the results for the kinetic term~(\protect\ref{kin2}).}
\label{fma}
\end{figure}
The data for the asymmetric state is also plotted in
figure~\ref{fma}. One can see good agreement between the Rayleigh-Ritz 
calculation and the Monte Carlo simulation with kinetic
term~(\ref{kin1}). From the data it seems however very unlikely that the results
for the different kinetic terms converge at $\epsilon=0$. In addition, a look at
the groundstate wave function (which can be obtained from histograms
in the MC calculation) shows a different behaviour for the two
kinetic terms as well.
This results comes as no surprise if one looks at a very simple case:
In~\cite{mic} it was shown that for $\vec{r}(t)$ and $\vec{r}(t+\epsilon)$ equal
but in
different patches, the kinetic
term~(\ref{kin1})
becomes $T_{kin} = (2\delta)^2(1-\delta/(2\pi))^2$, where
$\delta=\pi-r(t)$. The geodesic distance for this configuration is
exactly $(2\delta)^2$. If one follows terms of the form $\delta^3$
through the small-$\epsilon$ expansion, one finds they contribute
$O(\epsilon)$ to the external matrix element. But this exactly the
order which gives the zeroth order energy.

As mentioned earlier the geodesic distance is the solution of a 
quartic equation. Even for small $\epsilon$, configurations which
have multiple zeroes of the equation are not suppressed. The influence
of these configurations is beyond the scope of this article, but I
would suspect that they also contribute to the observed energy difference.

\section{An improved kinetic term}
Instead of using a kinetic term like~(\ref{kin1}), it is possible to use
a kinetic term which will exactly reproduce the Hamiltonian result
for the toy model. Again one would be free to tailor the kinetic term
to any boundary condition~(\ref{sy}). In order to make connection to
the Rayleigh-Ritz results of~\cite{baal}, I choose $b=0$. I make the Ansatz:
  \be e^{-\epsilon L_{kin}(\vec{r}_1,\vec{r}_2,\Delta k)} = \sqrt{2\pi
\epsilon}^3 \sum_{s=0}^1
\sum_{j,l=0}^\infty
(-1)^{s\Delta k} j_l(\alpha_{slj} r_1) j_l(\alpha_{slj} r_2)  P_l(c)
e^{-\frac12 \alpha_{s,j}^2 \epsilon},
\label{nkin}\ee 
where $c$ is the angle between $\vec{r}_1$ and $\vec{r}_2$ and
$\alpha_{0lj}$ and $\alpha_{1lj}$ are chosen according to~(\ref{sy}) and~(\ref{asy})
respectively. For large enough $\epsilon$ one can stop the sum at some values
$l_{max}$ and $j_{max}$ without introducing a large error (compared to the
statistical error of the MC-calculation).
For $\epsilon$ down to 0.15, I found
$l_{max}= 100$ and $j_{max}= 40$ to be sufficient.
 Of course it is
impracticable
to calculate the sum at every update step of the Monte Carlo
procedure. But one can tabulate the values and look them up during the
actual calculation. To this end we notice that~(\ref{nkin}) depends on
three continuous coordinates $r_1$,$r_2$ and $c$.
The influence from the boundary will
be large only for $\vec{r}_1$ and $\vec{r}_2$ close the boundary.
Otherwise
\be -\epsilon L_{kin}(\vec{r}_1,\vec{r}_2,0) \approx 
-\frac{1}{2\epsilon}(r_1^2+r_2^2-2r_1r_2 c),\label{beh1}\ee
and configurations with $\Delta k=1$ and $\vec{r}_1$ or $\vec{r}_2$ away
from the boundary are strongly suppressed.
Therefore I chose to tabulate $L_{kin}$ as a cubic spline interpolation
$S(r_1,r_2,c)$ in 3
dimensions, which is capable of describing~(\ref{beh1}) exactly.
For spline interpolation the gridpoints do not have to be equally
spaced (for a discussion of spline interpolation see for
example~\cite{numrec}).
A rough estimate suggest that the influence of the boundary,
i.e. the deviation from the quadratic form~(\ref{beh1}) is of the
order of $\exp[-(\pi-r)^2/(2\epsilon)]$. Therefore the gridpoints (in both r directions)
are chosen according to the distribution function
\be F(r)=\beta r + \rm{erf}(\pi/\sqrt{2\epsilon})-\rm{erf}((\pi-r)/\sqrt{2\epsilon}),\ee
where $\beta$ is chosen to minimize the maximum error\footnote{if $\beta$ is too
small
the interpolation starts to oscillate away from the boundary}.
For the spline interpolation one needs to store for each gridpoint eight
values:
\be S,
\frac{\partial^2 S}{\partial r_1^2},\frac{\partial^2 S}{\partial r_2^2},\frac{\partial^2 
S}{\partial c^2},
\frac{\partial^4 S}{\partial r_1^2 \partial r_2^2},\frac{\partial^4 S}{\partial r_1^2 \partial 
c^2},
\frac{\partial^4 S}{\partial r_2^2 \partial c^2},
\frac{\partial^6 S}{\partial r_1^2 \partial r_2^2 \partial
c^2}.\label{tabv} \ee
The eight quantities~(\ref{tabv}) can be calculated
from~(\ref{nkin}). (The derivatives at the
boundary which are needed to calculate~(\ref{tabv}) can also be
obtained from~\eq{nkin}). Once the spline table is set up the
calculation of the kinetic energy takes about 200 floating point
operations.

\begin{figure}
\begin{center}
\begin{picture}(10.0,7.0)
\put(0.5,6.0){$r$}
\put(3.6,-0.3){\it{support point}}
\includegraphics{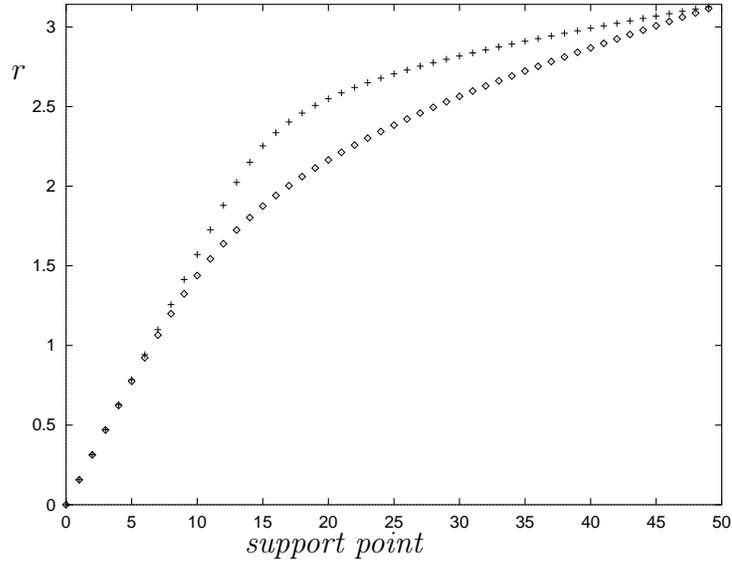}
\end{picture}
\end{center}
\caption{Location of support points for spline interpolation. Shown are
the locations for $\epsilon=0.6$ (diamonds) and $\epsilon=0.15$
(crosses).\label{grid}} 
\end{figure}

\begin{figure}
\epsfxsize=10cm
\centerline{\epsfbox{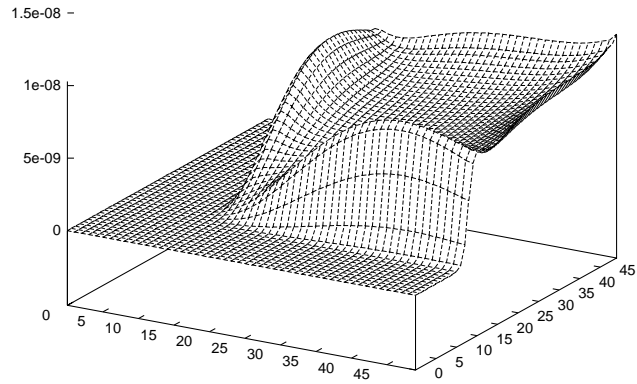}}
\caption{Absolute error of spline interpolation to the Boltzmann weight
 for both coordinates in the same sphere}
\label{er1}
\end{figure}

\begin{figure}
\epsfxsize=10cm
\centerline{\epsfbox{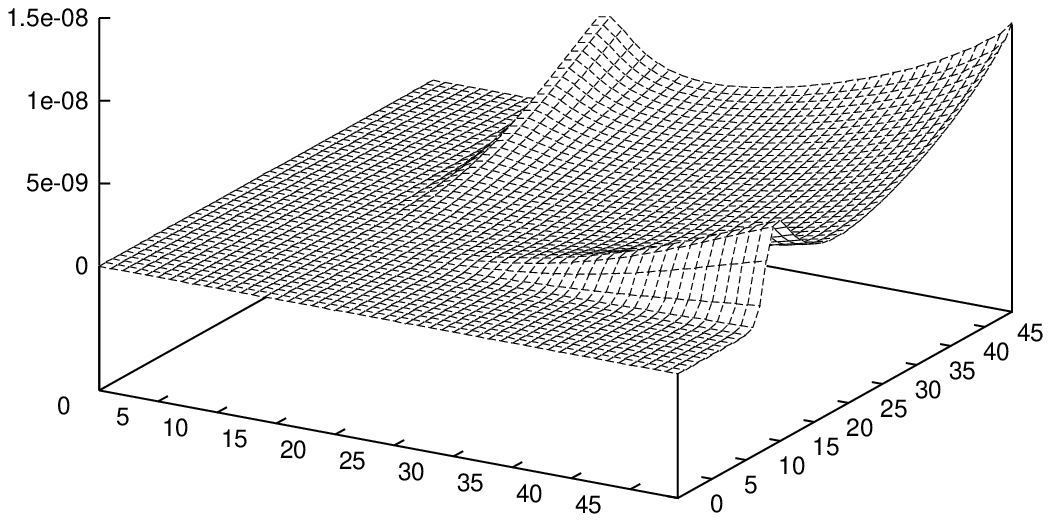}}
\caption{Absolute error of spline interpolation to the Boltzmann weight 
for the coordinates in different spheres}\label{er2} 
\end{figure}

For the simulations I used a $50^3$ grid. Due to the symmetry under
$r_1 \leftrightarrow r_2$ the size of the table is $8\times 2\times 50^3/2$ numbers, which
translates to 8 Mbytes when using IEEE double precision.
The placement of the r-gridpoints are shown in
figure~\ref{grid}.
In figures~\ref{er1} and~\ref{er2} an estimate of the error 
$| \exp[-\epsilon S]-\exp[-\epsilon L]|$ of the spline approximation
to the Boltzmann weight is shown.

For all $\epsilon$ used in the simulation the maximum absolute error
was about
$10^{-8}$ and can be neglected against the statistical error of the simulation. Due to
the Ansatz~(\ref{nkin}) the Monte Carlo calculation of the free particle
should reproduce the Hamiltonian results for any value of $\epsilon$.
In table~(\ref{mci}) data for $\epsilon=0.6$ is given. Smaller
$\epsilon$ give the same results within errors, but need more CPU-time
to obtain the same accuracy.
\begin{table}
\begin{center}
\caption{Monte Carlo results using the improved kinetic term.}\label{mci}
\vspace{.25cm}
\begin{tabular}{|l|l|l|}
\hline
state & exact & MC  \\ \hline
$A_1$ & 1.0 & 0.9985(15) \\ \hline
$A_2$ & 3.0 & 2.998(4)   \\ \hline
$\epsilon$  & 0.375 & 0.373(2) \\ \hline
$p$      & 0.25637.. & 0.2563(6)  \\ \hline
\end{tabular}\\
\end{center}
\end{table}

However, when one returns to the original problem there will be
$O(\epsilon^2)$ corrections, due to the potential energy term, and one has
to estimate the finite time step errors.
It is therefore necessary to simulate the Lagrangian at different
time steps $\epsilon$. To clarify the notation used for the results
let me write the Lagrangian in its original form:
\be \tilde{\epsilon} L = \half(g^{-2}+\alpha_1) \frac{(\Delta \vec{r})^2}{\tilde{\epsilon}}
+ \tilde{\epsilon} V(\vec{r}).\label{lorg}\ee
One can connect this form to a free particle and eq.~(\ref{nkin}) by
writing the kinetic term as 
\be \tilde{\epsilon} T_{kin} = \half\frac{(\Delta \vec{r})^2}{a}.\ee
This gives for the effective time step
\be \tilde{\epsilon} = a\left(\frac{1}{g^2}+\alpha_1\right),\ee which
has to be used to calculate the mass gap from the correlation
function. The kinetic energy term, however, has to be computed by using $a$ for
the time step in~\eq{nkin}.
Monte Carlo simulations were performed for $a=0.15,0.3,0.6$ and
$g=1.4,1.8,2.0$. The
results for various states are given in table~(\ref{mcr}), which also
contains an extrapolation down to $a=0$. One finds
excellent agreement for most states, indicating that the discrepancy
of the previously published results is entirely due to the kinetic term.
\begin{table}
\begin{center}
\caption{Masses for SU(2) small volume calculation. The values for
$a=0.15,0.3,0.6$ are measurements. The $a=0$ column is the extrapolation
to $a=0$ assuming an $a^2$ dependence of the mass. Cont. is the result
from the Hamiltonian approach~\protect\cite{baal}. For some states I was
unable to extract the mass for $a=0.15$, the corresponding entries are
left blank. $\epsilon, \epsilon_2$ and $\epsilon_3$ are states with 1, 2
and 3 units of electric flux. $A_1^+,E^+,T_2^+,A_1^-$ are IR of the
cubic group, $a_1^-$ is a $A_1^-$ IR with 3 units of electric flux and
$b_2^+$ is a $B_2^+$ representation of ${\cal{D}}_{4h}$ with one unit of
electric flux.\label{mcr}} 
\begin{tabular}{|r||l|l|l||l|l|}\hline
\multicolumn{6}{|c|}{$g=1.4$}\\
\hline
 state & a=0.6 & a = 0.3 & a = 0.15 & a = 0 & cont. \\ \hline
 $\epsilon$ & 0.2270(3) & 0.2034(6) & 0.1973(10) & 0.1954(6) & 0.195 \\ \hline
 $\epsilon_2$ & 0.480(3) & 0.430(5) & 0.418(3)   & 0.414(2)  & 0.416 \\ \hline
 $\epsilon_3$ & 0.787(6) & 0.709(9) & 0.689(10)  & 0.683(8)  & 0.686 \\ \hline
 $A_1^+$      & 1.35(6)  & 1.40(5)  & 1.52(3)    & 1.51(3)  & 1.509 \\ \hline
 $E^+  $      & 1.323(6) & 1.337(10)& 1.363(10)  & 1.356(8) & 1.354\\ \hline
 $T_2^+$      & 2.552(15)& 2.52(3)  &            & 2.51(4)  & 2.523 \\ \hline
 $b_2^+$      & 0.859(3) & 0.950(10)& 0.999(7)   & 1.002(7) & 1.007 \\ \hline
 $A_1^-$      & 5.42(15) & 5.5(2) & 5.52(5)      & 5.53(5)  & 5.479 \\ \hline
 $a_1^-$      & 6.08(10) & 6.1(2) &              & 6.1(3)   & \\ \hline
\multicolumn{6}{|c|}{$g = 1.8$} \\
\hline
 state & a=0.6 & a = 0.3 & a = 0.15 & a = 0 & cont. \\
\hline
$\epsilon$   & 0.523(1) & 0.507(2) & 0.506(3)  & 0.503(2) & 0.503 \\ \hline
$\epsilon_2$ & 1.133(4) & 1.103(6) & 1.098(15) & 1.094(7) & 1.095 \\ \hline
$\epsilon_3$ & 1.92(2)  & 1.86(3)  & 1.86(3)   & 1.85(3)  & 1.859 \\ \hline
$A_1^+$      & 2.43(5)  & 2.50(5)  & 2.56(10)  & 2.54(6)  & 2.509 \\ \hline
$E^+$        & 2.27(1)  & 2.254(10)& 2.27(2)   & 2.255(10)& 2.258 \\ \hline
$T_2^+$      & 4.46(4)  & 4.39(5)  & 4.35(7)   & 4.36(5)  & 4.350 \\ \hline
$b_2^+$      & 1.338(5) & 1.380(5) & 1.388(10) & 1.393(6) & 1.386 \\ \hline
$A_1^-$      & 5.92(4)  & 5.98(3)  & 6.02(4)   & 6.01(3)  & 5.987 \\ \hline
$a_1^-$      & 8.6(2)   & 8.4(1)   & 8.3(2)    & 8.3(2)   & \\ \hline
\multicolumn{6}{|c|}{$g = 2.0$} \\
\hline
 state & a=0.6 & a = 0.3 & a = 0.15 & a = 0 & cont. \\
\hline
$\epsilon$   & 0.7442(10) & 0.7312(15) & 0.727(3) & 0.727(3) & 0.728 \\ \hline
$\epsilon_2$ & 1.621(7)   & 1.588(10)  & 1.577(10)& 1.575(8) & 1.586 \\ \hline
$\epsilon_3$ & 2.72(2)    & 2.69(3)    & 2.67(3)  & 2.67(3)  & 2.672 \\ \hline
$A_1^+$      & 3.15(15)   & 3.30(10)   & 3.1(2)   & 3.27(15) & 3.177 \\ \hline
$E^+$        & 2.879(10)  & 2.884(15)  & 2.84(3)  & 2.87(2)  & 2.864 \\ \hline
$T_2^+$      & 5.54(10)   & 5.56(5)    & 5.49(7)  & 5.53(5)  & 5.513 \\ \hline
$b_2^+$      & 1.642(5)   & 1.657(10)  & 1.62(3)  & 1.656(15)& 1.660 \\ \hline
$A_1^-$      & 6.05(2)    & 6.08(3)    &          & 6.09(4)  & 6.113 \\ \hline
$a_1^+$      & 10.2(3)    & 10.3(2)    & 10.1(2)  & 10.2(2)  & 9.79\\ \hline
\end{tabular}
\end{center}
\end{table}

\section{Conclusions}
The treatment of non-trivially connected coordinate spaces using 
Monte Carlo methods needs some care. There are basically two effects
which hinder the MC approach with a naive kinetic term.
Firstly one encounters corrections of the order of the square root of
the time step, which require very small time steps and therefore large
amounts of CPU time. Secondly, and more
severe the conditions at the coordinate patch connection are not under
control. Usually one would argue that the Monte Carlo approach will
choose the correct boundary conditions. But if the connection of the
patches is not simple, this argument is on very weak footing, and
the boundary condition (and spectrum) preferred by the simulation depends on
details of the kinetic term.
The problem
can be overcome by constructing a kinetic term which will match
predetermined conditions at the boundary. In the case studied here,
the term could be tabulated and used efficiently in a MC calculation.
The improved kinetic term has no finite time step
correction, and one is left with $O(\epsilon^2)$ correction from the
potential energy.

Using the improved kinetic term,
I was able to reproduce the Rayleigh-Ritz results
of~\cite{baal} with the Lagrangian approach. Note that both
approaches have the same degree of freedom in choosing the boundary
conditions.
In the original Lagrangian approach~\cite{mic} the freedom of choice
is not as obvious, because it is hidden non-trivially in the choice of
sub-leading orders of the kinetic term.

Let me remark at this point that the transfer matrix approach
of~\cite{baal5} is unrelated to the approach presented here, that is
van Baal's result of small finite time step corrections does not
contradict my findings.

\section*{Acknowledgements}
This research project was partially
funded by the Department of Energy under contracts
DE-FG05-87ER40319 and DE-FC05-85ER2500.
Finally, I would like to thank B. Berg for comments on the manuscript.

\end{document}